\documentclass[unnumsec,webpdf,contemporary,large]{oup-authoring-template}
\graphicspath{{figures/}}

\usepackage{bbding}
\DeclareRobustCommand{\FourStarSmall}{%
  \raisebox{0.10ex}{\scalebox{0.65}{\mbox{\FourStar}}}%
}
\setcitestyle{sort&compress,super,comma,open={},close={}}
\begin{document}

\journaltitle{Journal Title Here}
\DOI{DOI added during production}
\copyrightyear{YEAR}
\pubyear{YEAR}
\vol{XX}
\issue{x}
\access{Published: Date added during production}
\appnotes{Paper}

\firstpage{1}
\title[]{Structurally Constrained Brain Network Dynamics Reveal Reduced Functional Flexibility in Cocaine Use Disorder}

\author[1]{Seyed Majid Razavi}
\author[2]{Saeed Tajik Hesarkuchak}
\author[3]{Triet M. Tran}
\author[2]{Mehdi Zaeifi}
\author[2]{Amirhossein Arezoumand}
\author[1,4]{Farnaz Zamani Esfahlani}
\author[5,6]{Jason A. Oliver}
\author[1,2$\ast$]{Sina Khanmohammadi}

\address[1]{%
\orgdiv{Data Science and Analytics Institute},
\orgname{University of Oklahoma},
\orgaddress{\street{202 W. Boyd St.}, Norman,
\postcode{73019}, \state{OK}, \country{USA}}}

\address[2]{%
\orgdiv{School of Computer Science},
\orgname{University of Oklahoma},
\orgaddress{\street{110 W. Boyd St.}, Norman,
\postcode{73019}, \state{OK}, \country{USA}}}

\address[3]{%
\orgdiv{Brain and Spine Institute},
\orgname{University of Tennessee Medical Center},
\orgaddress{\street{1928 Alcoa Hwy.}, Knoxville,
\postcode{37920}, \state{TN}, \country{USA}}}

\address[4]{%
\orgdiv{Stephenson School of Biomedical Engineering},
\orgname{University of Oklahoma},
\orgaddress{\street{173 Felgar St.}, Norman,
\postcode{73019}, \state{OK}, \country{USA}}}

\address[5]{%
\orgdiv{Department of Family and Preventive Medicine},
\orgname{University of Oklahoma Health Campus},
\orgaddress{\street{655 Research Parkway}, Oklahoma City,
\postcode{73104}, \state{OK}, \country{USA}}}

\address[6]{%
\orgdiv{TSET Health Promotion Research Center},
\orgname{University of Oklahoma Health Campus},
\orgaddress{\street{655 Research Parkway}, Oklahoma City,
\postcode{73104}, \state{OK}, \country{USA}}}

\corresp[$\ast$]{Corresponding author. \href{email:sinakhan@ou.edu}{sinakhan@ou.edu}}

\received{Date}{0}{Year}
\revised{Date}{0}{Year}
\accepted{Date}{0}{Year}

\abstract{Cocaine Use Disorder (CUD) is associated with widespread alterations in large-scale functional brain networks, yet the mechanisms contributing to these changes and their relationship to clinical and cognitive outcomes remain poorly understood. To address this gap, we introduce a framework to extract structurally informed dynamic functional connectivity patterns. We then leverage these connectivity patterns to characterize differences in functional brain network organization associated with CUD and to examine their relationship with clinical measures. Specifically, we applied Laplacian spectral smoothing to each participant's functional connectivity matrix using individualized structural priors derived from diffusion imaging. These structurally informed connectivity features were subsequently used to examine cross-network interactions and characterize dynamic community organization across functional brain states. Our findings indicate that individuals with cocaine use disorder exhibit increased integration and recruitment accompanied by reduced flexibility in the functional brain networks, with the most pronounced alterations in visual, attentional, and control systems. In addition, structurally informed functional connectivity features were predictive of weekly cocaine use within the CUD cohort. Overall, these results highlight the value of structurally informed dynamic connectivity measures for characterizing network-level alterations associated with cocaine addiction and for linking these alterations to clinically meaningful measures of cocaine use severity.}

\keywords{Cocaine Use Disorder, Resting-state fMRI, Diffusion MRI, Structurally Informed Functional Connectivity, Dynamic Multilayer Networks,
Community Detection}

\maketitle
\section{Introduction}
Cocaine Use Disorder (CUD) is characterized by intense craving, compulsive drug seeking, and lack of control over substance use \citep{koob2008role,dalley2011impulsivity}. These features are associated with impaired emotion regulation, attention deficits, and reduced response inhibition \citep{kubler2005cocaine, goldstein2011dysfunction, koob2016neurobiology}. These disturbances contribute to the persistence of addictive behavior by weakening decision-making capacity and increasing vulnerability to relapse \citep{volkow2010addiction,everitt2016drug}. Therefore, a deeper understanding of the neurobiological mechanisms underlying cocaine addiction is essential for developing targeted interventions that not only curb drug-seeking behavior but also remediate the cognitive and emotional dysregulation that sustains addiction.

Neuroimaging studies have linked CUD to widespread alterations in brain structure, including reductions in gray matter volume \citep{dang2022meta}, white matter integrity \citep{narayana2014chronic}, and decreased cortical thickness, particularly in prefrontal, temporoparietal, and cingulate regions \citep{schinz2023lower}. Further research has found that these structural changes are linked to specific factors, such as length of cocaine use \citep{ersche2011abnormal}, genotype \citep{alia2011gene}, and impulsivity \citep{moreno2012trait}. However, these structural changes are not limited to gray matter; they also affect the white matter connections between different regions of the brain. Diffusion Magnetic Resonance Imaging (dMRI) studies showed that white matter integrity is compromised in cocaine users, including in major tracts such as the corpus callosum \citep{moeller2005reduced, suchting2021meta}. Several studies have also pointed to significant disruptions in functional connectivity within and between large-scale brain networks in cocaine use disorder, including the default mode network (DMN) \citep{ding2013cocaine}, salience network (SN) \citep{li2024brain}, and central executive network (CEN) \citep{woisard2023executive}. These disruptions have been associated with alexithymia \citep{liang2015interactions} and diminished top-down cognitive control \citep{worhunsky2013functional}.

Despite growing evidence of structural and functional disruptions in cocaine addiction, most of the prior studies have examined these alterations in isolation, treating brain structure and functional connectivity as independent constructs \citep{rasgado2024structural}. In the present work, we address this limitation by integrating individual structural connectivity (SC) metrics as an explicit mathematical prior to constrain time‑resolved functional connectivity estimates. This approach incorporates underlying anatomical architecture into the estimation of dynamic Functional Connectivity (FC), enabling a more precise characterization of connectivity patterns associated with cocaine addiction. Specifically, we used community organization metrics derived from these structurally informed functional connectivity (SiFC) networks to examine both local and global alterations in network dynamics associated with cocaine use disorder, evaluate their ability to distinguish individuals with CUD from Healthy Controls (HC), and assess their relationship with weekly cocaine use.

\section{Materials and Methods}
\subsection{Data Description}
\subsubsection{Participants:} 
In this study, we used the SUDMEX-CONN dataset \citep{angeles2022mexican}, an open-access dataset from individuals with cocaine use disorder and healthy controls. The dataset includes demographic, clinical, cognitive, and imaging data. The CUD status was established using the Spanish version 5.0.0 of the Mini International Neuropsychiatric Interview–Plus (MINI-Plus) \citep{sheehan1998mini}. The summary of demographic and clinical information for both groups is presented in Table \ref{tab1}, where no significant differences were found between the groups in terms of age ($p = 0.51$), sex ($p = 0.47$), or years of education ($p = 0.33$). The $p$-values were computed using two-sample t-tests for continuous variables and a $\chi^2$ test for categorical ones. The weekly cocaine dosage (g/week) was also recorded for the CUD group. 

\begin{table}
\caption{Demographic and Clinical Characteristics of Participants\label{tab1}}%
\footnotesize
\setlength\tabcolsep{2pt}
\renewcommand{\arraystretch}{1.75} 
\begin{tabular*}{\columnwidth}{@{\extracolsep\fill}llll@{\extracolsep\fill}}
\toprule
               & Cocaine use & Healthy & Statistics \\[-1ex]
Characteristic & disorder    & controls & ($p$-value) \\
\midrule
Age (years)        & 31.92 $\pm$ 8.03   & 30.67 $\pm$ 8.24   & 0.51 \\
Sex (Male/Female)  & 33/5               & 30/8               & 0.47 \\
Education (years)  & 11.54 $\pm$ 3.01   & 12.26 $\pm$ 3.14   & 0.33 \\
Weekly Dose (g/week)    & 3.09 $\pm$ 1.22    &  ---                 &  ---  \\
\botrule
\end{tabular*}
\begin{tablenotes}%
\item * $p$-values were computed using two-sample t-tests for continuous variables and a $\chi^2$ test for categorical variables. 
\end{tablenotes}
\end{table}

\subsubsection{Image Acquisition:}
Magnetic Resonance Imaging (MRI) data were acquired using a 3T Philips Ingenia MRI scanner equipped with a 32-channel dS Head Coil. The imaging protocol included three modalities: structural T1-weighted MRI, multishell high-angular-resolution diffusion-weighted imaging (HARDI-DWI), and resting-state functional MRI (rs-fMRI). Structural images were acquired using a three-dimensional fast field echo (3D-FFE) SENSE sequence with the following parameters: repetition time (TR) = 7 ms, echo time (TE) = 3.5 ms, field of view (FOV) = $240 \times 240~\mathrm{mm}^2$, and isotropic voxel size = $1 \times 1 \times 1~\mathrm{mm}^3$. Resting-state fMRI data were collected using a gradient-echo echo-planar imaging (GE-EPI) sequence with TR/TE = $2000/30$ ms, flip angle = $75^{\circ}$, voxel size = $3 \times 3 \times 3~\mathrm{mm}^3$, and 36 axial slices. During the resting-state scan, participants were instructed to keep their eyes open, fixate on a cross displayed on a screen, remain still, and stay awake. Diffusion-weighted images were acquired using a spin-echo echo-planar imaging (SE-EPI) sequence with multishell HARDI sampling. Acquisition parameters included TR/TE = \(8600/126.78\) ms, isotropic voxel size = $2 \times 2 \times 2~\mathrm{mm}^3$, and 136 diffusion volumes distributed across two non-zero $b$-value shells. Specifically, the protocol included 8 non-diffusion-weighted (\(b = 0\)) volumes, 36 diffusion directions with $b = 1000~\mathrm{s/mm^2}$, and 92 diffusion directions with $b = 3000~\mathrm{s/mm^2}$.

\subsection{Structurally Informed Multilayer Functional Connectivity} 
\begin{figure*}
    \centering
   \includegraphics[height=0.8\textheight, keepaspectratio]{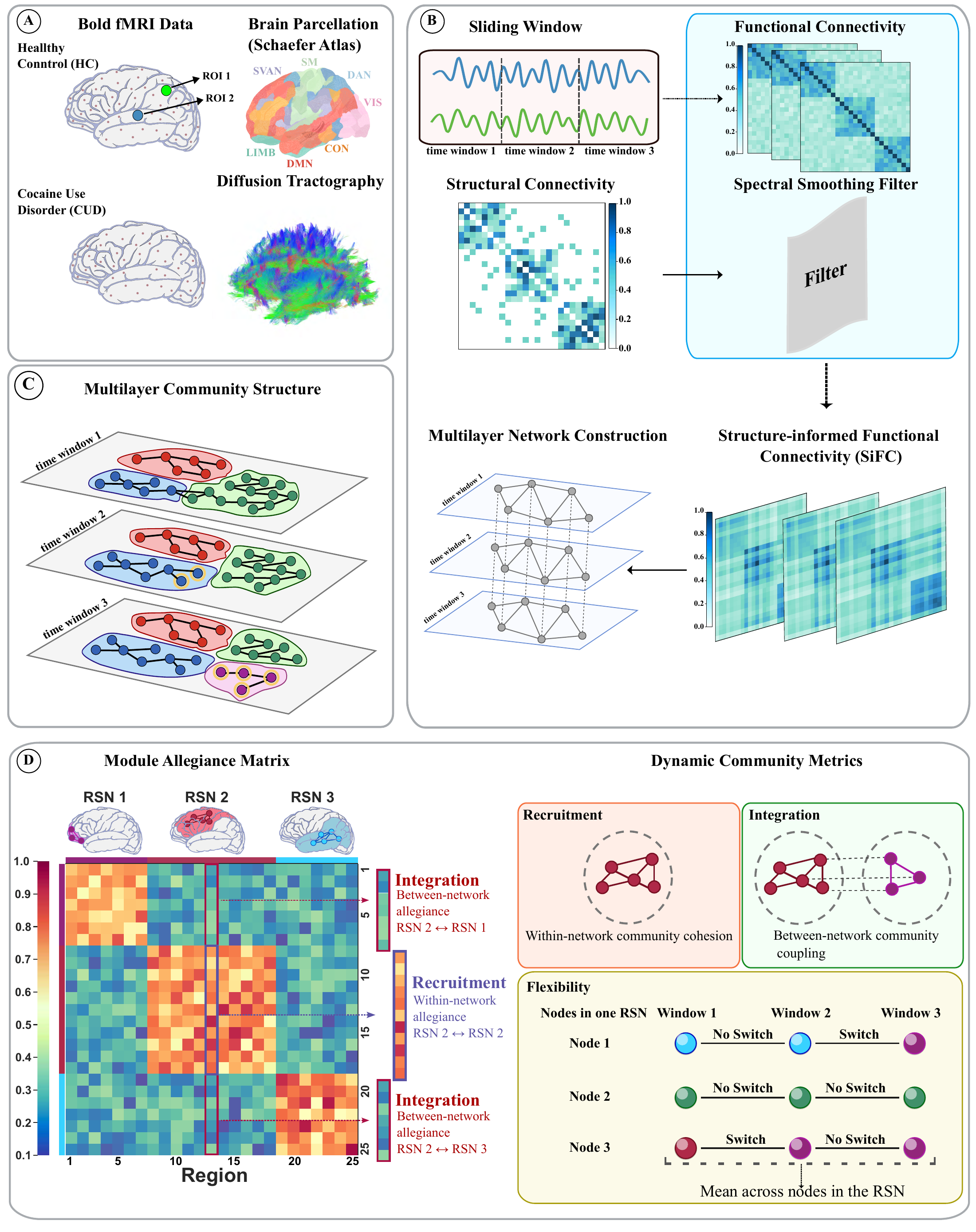}
    \caption{\textbf{Overview of the structurally informed temporal multilayer functional connectivity framework.}
\textbf{(A) Data Preparation.} Resting-state fMRI time series were extracted from $N=200$ cortical regions (Schaefer atlas) for 38 healthy controls and 38 cocaine use disorder participants. \textbf{(B) Network Construction.} The preprocessed fMRI time series were segmented into non-overlapping temporal windows to compute windowed functional connectivity matrices using Pearson correlation.  Diffusion-weighted imaging data were processed to generate whole-brain tractography and subject-specific structural connectivity matrices. A subject-specific spectral smoothing filter was then derived from the normalized graph Laplacian of structural connectivity matrices and applied to the functional connectivity matrix in each window to obtain structurally informed functional connectivity matrices. These SiFC layers were stacked to form a temporal multilayer network, with ordinal coupling that links each node to itself across adjacent windows. \textbf{(C) Multilayer Community Detection.} Generalized multilayer modularity maximization was used to identify communities across time, producing window-specific community assignments. \textbf{(D) Analysis of Community Dynamics.} A module allegiance matrix was computed, representing the probability that two regions are assigned to the same community across windows (and repetitions). From the module allegiance matrix, we derived recruitment (within-network co-assignment for each canonical network) and integration (between-network co-assignment with other systems) to summarize dynamic community organization across the seven canonical networks from Yeo-Krienen atlas. Flexibility was also computed as the frequency with which a node changed its community assignment between consecutive windows.} \label{fig:framework}
\end{figure*}

The overall framework of structurally informed multilayer functional connectivity analysis is shown in Fig.~\ref{fig:framework}. The four main steps include data preparation, network construction, multilayer community detection, and analysis of community dynamics. Each step is described in detail below.

\subsubsection{Data Preparation}
\paragraph{Image Preprocessing:}
Structural and functional MRI data from 76 participants, including 38 healthy controls and 38 individuals with cocaine use disorder, were preprocessed using fMRIPrep version 23.2.1 \citep{esteban2019fmriprep}. Anatomical T1-weighted (T1w) images underwent skull-stripping, bias field correction, and registration to the standard MNI152NLin2009cAsym space using a combination of FreeSurfer \citep{fischl2012freesurfer} and ANTs registration \citep{avants2011reproducible}. Functional MRI data preprocessing included removal of the initial two volumes to account for magnetization equilibrium, followed by slice-timing correction based on the interleaved slice acquisition order, and rigid-body head motion correction to realign all volumes to the mean functional image. To address susceptibility-induced distortions, field maps acquired with opposite phase encoding directions were utilized. Functional images were coregistered to corresponding structural images and spatially normalized to the $2$~mm isotropic resolution MNI152NLin2009cAsym template.

The effects of confounding signals were reduced using nuisance regression, which included the 24-parameter Friston head-motion model \citep{friston1996movement}, Cerebrospinal Fluid (CSF) and White Matter (WM) signals, and anatomical CompCor components \citep{behzadi2007component}. We applied a high-pass temporal filter (cutoff 0.01~Hz) to remove low-frequency drifts. The initial two volumes were also removed from the confound regressors to maintain temporal consistency with the BOLD time series. Region-specific time series extraction for network analysis was conducted using the Schaefer atlas \citep{schaefer2018local}, comprising 200 cortical parcels grouped into seven Resting-State Networks (RSNs) including visual (VIS), somatomotor (SM), dorsal attention (DAN), salience/ventral attention (SVAN), limbic (LIMB), control (CON), and default mode (DMN) according to the Yeo-Krienen atlas \citep{yeo2011organization}, available at a 2mm resolution in MNI152NLin2009cAsym space.  

Region of Interest (ROI) time series were then extracted from unsmoothed functional data to avoid artificially inflating local correlations and to preserve the validity of subsequent network analyses. The resulting dataset consisted of detrended ROI time series from 200 cortical regions (298 time points per run), which were subsequently segmented using a moving window approach. It should be noted that, because the Schaefer atlas comprises only cortical parcels, all ROI-based analyses were restricted to cortical regions, with subcortical structures excluded from consideration. Consequently, labels such as the LIMB network refer to cortical parcels assigned to the limbic network in the Yeo-Krienen functional parcellation rather than the subcortical limbic structures.

Diffusion-weighted imaging data preprocessing was performed using QSIPrep version 0.18.1 \citep{cieslak2021qsiprep}. Raw diffusion images underwent denoising, eddy-current correction, and susceptibility-induced distortion correction using field maps with opposite phase encoding directions. Anatomical T1-weighted images were skull-stripped, bias-field corrected, and registered to native ACPC orientation. Subsequently, the processed DWI data were aligned to the native anatomical space (ACPC) and resampled to an isotropic voxel resolution of 2mm.

\paragraph{Quality Control:}
Quality control included visual inspection of anatomical and functional images, assessment of motion artifacts, and evaluation of motion parameters. Runs exhibiting excessive head motion, defined as framewise displacement (FD) exceeding $0.5$ mm in more than 20\% of volumes, were excluded. Additionally, participants lacking clear group labels (CUD or HC) were removed to maintain consistency and integrity across study groups. Quality control procedures for DWI data included visual inspection of diffusion and anatomical alignment, gradient direction consistency checks, and streamline distribution assessment. Participants with low DWI data quality or incomplete preprocessing were excluded to ensure data integrity. 

\subsubsection{Network Construction}
\paragraph{Structural Connectivity:}
Structural connectivity was derived using MRtrix3 \citep{tournier2019mrtrix3}. To ensure anatomical correspondence between modalities, we used the same Schaefer-200 parcellation for both fMRI time-series extraction and diffusion-based connectome construction. Initially, the Schaefer cortical atlas in MNI152NLin2009cAsym space was warped into each subject's native ACPC space using ANTs non-linear registration \citep{avants2008symmetric}. DWI data and associated gradient directions (bvec/bval) were converted into MRtrix-compatible formats. Anatomically Constrained Tractography (ACT) tissue priors were generated from the native-space T1-weighted images. Fiber orientation distributions (FODs) were estimated using multi-shell, multi-tissue constrained spherical deconvolution (MSMT-CSD) \citep{jeurissen2014multi}. Whole-brain probabilistic tractography was performed using the iFOD2 algorithm, generating 10 million streamlines per subject. To refine streamline distributions and reduce biases, streamline weights were optimized using SIFT2 \citep{smith2015sift2}. Finally, streamline counts weighted by SIFT2 were aggregated into SC matrices using region definitions from the Schaefer atlas ($200\times200$ cortical connectome), yielding a subject-specific $200\times200$ SC matrix for subsequent analysis. Structural connectivity matrices were treated as weighted and symmetric networks using SIFT2-weighted streamline estimates. No additional thresholding was applied prior to Laplacian construction.

\paragraph{Functional Connectivity:}
We segmented each subject's regional time series into 19 non-overlapping 30s windows by discarding the final 13 volumes, yielding $19 \times 15$ volumes of fMRI scans. A 30s window balances temporal resolution against estimation reliability for dynamic FC, consistent with prior guidance that suggests 30–60s windows for resting-state dynamics \citep{leonardi2015spurious, gifford2020resting}. For each window, we computed the $200 \times 200$ Pearson correlation matrix across regions and set the diagonal to zero to remove self-connections. We then retained only positive correlations to obtain a nonnegative weighted FC matrix. Positive and negative connections play fundamentally different roles in community detection, with positive edges promoting co-assignment of regions and negative edges promoting their separation into different communities. By focusing on positive functional interactions, the resulting communities can be interpreted directly as groups of regions exhibiting coordinated functional coupling over time. Stacking the $19$ window-level matrices yielded a temporal multilayer network with $L = 19$ layers per participant, where layers are ordered in time and inter-layer links are permitted only between the same region across consecutive layers (ordinal coupling). 

\paragraph{Structurally Informed Functional Connectivity:}
To incorporate anatomical constraints, we derived a subject-specific spectral filter from each participant's structural connectivity matrix 
$S$. We construct the normalized Laplacian from $S$ as:
\begin{equation}
\mathcal{L} \;=\; I - D^{-1/2}\,S\,D^{-1/2}.
\end{equation}
where, $D=\text{diag}(d_1,\dots,d_N)$ is the degree matrix with $d_i=\sum_j S_{ij}$. The eigendecomposition of $\mathcal{L}$ is given by $ \mathcal{L} = U \Lambda U^{\top} $, where $U$ is an orthonormal matrix whose columns are the eigenvectors of $\mathcal{L}$, and $\Lambda = \mathrm{diag}(\lambda_1, \dots, \lambda_N)$ is a diagonal matrix of nonnegative eigenvalues. We define a low-pass graph filter with spectral response $h(\lambda)=\frac{1}{1+\tau\lambda}$ for $\tau>0$ given by:  
\begin{equation}
G \;=\; U\,\text{diag}\!\bigl(h(\lambda_1),\dots,h(\lambda_N)\bigr)\,U^{\top}
\;=\; (I+\tau \mathcal{L} )^{-1}, 
\end{equation}

where, $\tau$ is the smoothing parameter. Compared with direct structural masking or element-wise multiplication by the SC matrix, $G$ provides a tunable isotropic structural graph regularizer rather than forcing functional connectivity to follow the structural network exactly. This allows the window-level FC edge-weight estimates to be smoothed according to the subject-specific structural connectome while still preserving functional coupling patterns estimated directly from the observed fMRI time series. For each FC layer $s \in \{1,\dots,L\},$ the structurally informed functional connectivity is obtained by bilateral graph smoothing followed by the same post-processing as in functional network construction (set the diagonal to zero and enforce nonnegativity). 
\begin{equation}
A'^{(s)} = G A^{(s)}G,
\end{equation}
The result is a subject-specific SiFC multilayer tensor for subsequent steps including multilayer community detection. The spectral filtering procedure can be viewed as an anatomically informed regularization strategy that smooths functional connectivity estimates according to the topology of the underlying structural connectome.

\subsubsection{Multilayer Community Detection}

\paragraph{Generalized Multilayer Modularity:}
Communities in a graph represent groups of nodes that are more highly connected to one another than to nodes outside of their community \citep{newman2006modularity}. To identify dynamic whole-brain and network-level community organization  in our temporal multilayer networks, we used the generalized multilayer modularity framework \citep{mucha2010community,canal2024dynamic}.

Let $A'^{(s)}\in\mathbb{R}^{N\times N}$ denote the structurally informed functional connectivity for layer $s\in \{1,\dots,L\}$ (symmetric, nonnegative, and with zero diagonal), and let $c_i^{(s)}$ and $c_j^{(r)}$ be the community labels of nodes $i$ and $j$ in layers $s$ and $r$, respectively. The generalized multilayer modularity is then defined as:
\begin{equation}
Q = \frac{1}{2\mu}\sum_{i,j=1}^{N}\sum_{s,r=1}^{L}
\Big[\big(A'^{(s)}_{ij}-\gamma P^{(s)}_{ij}\big)\delta_{sr}
+\delta_{ij}\omega\delta_{r,s\pm1}\Big]
\delta\big(c_i^{(s)},c_j^{(r)}\big),
\end{equation}

where $2\mu$ denotes the total edge weight of the multilayer network, $\gamma$ is the resolution parameter that rescales the intralayer contributions, and $P^{(s)}_{ij}$ is the Newman-Girvan null model for layer $s$. The symbols $\delta_{sr}$ and $\delta_{ij}$ are Kronecker delta functions (equal to $1$ when their indices are equal and $0$ otherwise). The parameter $\omega$ denotes the temporal (interlayer) coupling strength, $\delta_{r,s\pm1}$ restricts interlayer coupling to adjacent layers (i.e., $r=s\pm1$), and $\delta\big(c_i^{(s)},c_j^{(r)}\big)$ is an indicator function that equals $1$ if nodes $i$ and $j$ belong to the same community across layers $s$ and $r$, and $0$ otherwise. 

The Newman–Girvan null model for layer $s$ is defined as:
\begin{equation}
P^{(s)}_{ij}=\dfrac{k^{(s)}_i k^{(s)}_j}{2m^{(s)}},
\end{equation}

where $k^{(s)} = A'^{(s)} \mathbf{1}$ is the vector of node strengths in layer $s$ (i.e., $k^{(s)}_i = \sum_{j} A'^{(s)}_{ij}$), and $m^{(s)} = \tfrac{1}{2}\mathbf{1}^\top k^{(s)} = \tfrac{1}{2}\sum_{i,j} A'^{(s)}_{ij}$ is the total edge weight in layer $s$. The matrix $P^{(s)}_{ij}$ represents the Newman-Girvan configuration-model expectation, which preserves the node strength sequence within each layer while randomizing connections. Hence, it provides a baseline against which the observed connectivity $A'^{(s)}_{ij}$ is compared in the modularity function, allowing the identification of statistically significant community structure.

In our implementation to optimize the multilayer modularity model, we build a block-structured supra-adjacency matrix whose diagonal blocks capture within-window structure relative to a strength-preserving null model, and whose immediate off-diagonal blocks encode ordinal coupling between adjacent windows; all other blocks are zero \citep{mucha2010community}. We then maximize $Q$ with the generalized Louvain algorithm, and given that this process is stochastic, we performed 100 independent repetitions per subject with randomized initial conditions, retaining the community labels $C$ whose entries are $c_i^{(s)}$ for all node–layer pairs and the corresponding modularity values  $Q$  \citep{jutla2011generalized}.  Stacking across repetitions yields an array of $C$ and a vector of $Q$ that we use for downstream dynamic measures. All analysis parameters are kept constant across participants to ensure comparability. Because all layers are optimized jointly on the same supra-graph, the community assignments are directly comparable across windows \citep{mucha2010community}.

\paragraph{Parameter Settings:}
The resolution parameter $\gamma$ sets the weight of connections within each layer $s$ relative to their null expectation. By varying $\gamma$, we control the size and number of detected communities, where low $\gamma$ produces fewer but larger communities and high $\gamma$ produces more but smaller communities \citep{newman2004finding, newman2006modularity}. The interlayer coupling parameter $\omega$, on the other hand, governs the strength of identity links that connect each node to itself across layers. When $\omega$ is large, the identity links over time are stronger, and a community found in window $s$ tends to persist into windows $s-1$ and $s+1$. When $\omega$ is small, the optimization emphasizes within-window structure, and communities can reconfigure more freely over time. Hence, the coupling parameter $\omega$ controls how much a node is encouraged to keep the same community across consecutive windows \citep{puxeddu2020modular}. In our study, we set $\gamma=1$ and $\omega=0.5$ to explore an intermediate regime where modules can reconfigure over time while maintaining comparability across consecutive windows.  We also set the structural smoothing parameter to $\tau=0.3$, resulting in a moderate level of anatomical constraint that preserves underlying structural organization without overly smoothing functional dynamics. To evaluate the robustness of our results with respect to parameter selection, we conducted a sensitivity analysis across multiple parameter settings. The results are presented in the supplementary material.

\subsubsection{Analysis of Community Dynamics}
\paragraph{Measures of Dynamic Community Structure:}

We characterize dynamic community organization across temporal layers using three measures derived from the multilayer community assignments \citep{canal2024dynamic}. First, we compute the module allegiance (MA) matrix, which quantifies the probability that pairs of regions are assigned to the same community across temporal layers and independent Louvain runs. Let $c_i^{(s,o)}$ denote the community assignment of region $i$ in temporal layer $s$ obtained from Louvain run $o$, where $L$ and $O$ denote the numbers of temporal layers and independent Louvain runs, respectively. The module allegiance between regions $i$ and $j$ is defined as
\begin{equation}
\mathrm{MA}_{ij} = \frac{1}{OL} \sum_{o=1}^{O} \sum_{s=1}^{L} \delta\!\left( c_i^{(s,o)}, c_j^{(s,o)} \right),
\end{equation}

where $\delta(\cdot,\cdot)$ is an indicator function that equals $1$ if two regions are assigned to the same community and $0$ otherwise. Thus, $\mathrm{MA}_{ij}\in[0,1]$ represents the fraction of layer-run combinations in which regions $i$ and $j$ are assigned to the same community. From the MA matrix, we derive two network-level summaries for the seven canonical networks: (1) Recruitment, which is the within-network likelihood that regions of the same canonical network are assigned to the same community across temporal layers, reflecting internal cohesion, and (2) Integration, which is the across-network likelihood that regions of a given network are assigned to the same community as regions from other networks across temporal layers, summarizing cross-system coupling \citep{bassett2015learning, mattar2015functional}. 

Let $\mathcal{V}_a$ denote the set of regions belonging to canonical network $a$. For a region $i \in \mathcal{V}_a$, regional recruitment and integration are defined as:

\begin{align} \mathrm{Rec}_{i}^{(a)} &= \frac{1}{|\mathcal{V}_a|-1} \sum_{\substack{j\in\mathcal{V}_a\\j\neq i}} \mathrm{MA}_{ij}, \\ \mathrm{Int}_{i}^{(a)} &= \frac{1}{N-|\mathcal{V}_a|} \sum_{j\notin\mathcal{V}_a} \mathrm{MA}_{ij}. \end{align} 

Here, $\mathrm{Rec}_{i}^{(a)}$ quantifies the average module allegiance between region $i$ and all other regions within its canonical network, whereas $\mathrm{Int}_{i}^{(a)}$ quantifies the average module allegiance between region $i$ and regions outside its canonical network. Network-level recruitment and integration are then computed by averaging the corresponding regional measures across all regions in canonical network $a$: 

\begin{align} \mathrm{Rec}_{a} &= \frac{1}{|\mathcal{V}_a|} \sum_{i\in\mathcal{V}_a} \mathrm{Rec}_{i}^{(a)}, \\ \mathrm{Int}_{a} &= \frac{1}{|\mathcal{V}_a|} \sum_{i\in\mathcal{V}_a} \mathrm{Int}_{i}^{(a)}. \end{align}

Next, we quantify flexibility, which measures how frequently a brain region changes its community between consecutive temporal layers. The flexibility of region $i$ is defined as:

\begin{equation} \mathrm{Flex}_{i} = \frac{1}{O(L-1)} \sum_{o=1}^{O} \sum_{s=1}^{L-1} \left[ 1-\delta\!\left( c_i^{(s,o)}, c_i^{(s+1,o)} \right) \right]. \end{equation}

Thus, $\mathrm{Flex}_{i}\in[0,1]$ quantifies the tendency of region $i$ to change its community assignment between consecutive temporal layers. In addition to reporting flexibility averaged within each of the seven canonical Yeo--Krienen networks (VIS, SM, DAN, SVAN, LIMB, CON, and DMN), we also report node-wise flexibility across the whole brain to characterize the spatial distribution of community reconfiguration across the cortex \citep{bassett2015learning,pedersen2018multilayer}. Network-level flexibility for canonical network $a$ is computed by averaging the node-wise flexibility values across all regions in that network: 

\begin{equation} \mathrm{Flex}_{a} = \frac{1}{|\mathcal{V}_a|} \sum_{i\in\mathcal{V}_a} \mathrm{Flex}_{i}. \end{equation}

\paragraph{Statistical Analysis:}

To assess the statistical significance of group differences (HC vs CUD) in resting-state network dynamics (flexibility, recruitment, and integration), we used two-sided nonparametric permutation tests with $10000$ permutations \citep{bassett2008hierarchical, he2008structural}. For each RSN and metric, we computed the observed difference in group medians and compared it against a null distribution obtained by randomly permuting group labels. Prior to testing, we removed outliers using the median absolute deviation (MAD) rule, where outliers within each group were defined as values exceeding 3 MAD from that group's median. To control for multiple comparisons, we applied false discovery rate (Benjamini–Hochberg FDR) corrections at $q<0.05$ across all tests, including RSN-level contrasts (3 measures $\times$ 7 networks) and whole-brain contrasts (3 measures) \citep{benjamini1995controlling}.

\paragraph{Predictive Analysis:}
To evaluate whether dynamic network measures contain information that differentiates individuals with CUD from HC, we trained logistic regression classifiers using the 21 RSN-level predictors (flexibility, recruitment, and integration across seven networks). We compared two models, an FC  model using conventional functional connectivity features and a SiFC model that utilizes the structurally informed connectivity features. A stepwise feature selection method based on Akaike Information Criterion (AIC) was utilized to identify the most relevant features for each model. The trained models were evaluated with fixed stratified 10-fold cross-validation, where at each fold two performance metrics of Receiver Operating Characteristic (ROC) and the Area Under the Curve (AUC) were calculated. Uncertainty in AUC was summarized using 95\% nonparametric bootstrap confidence intervals $(B_{\text{boot}}=5000)$ based on subject-level resampling of cross-validated predictions. For a single operating point summary, we selected the decision threshold that maximized Youden's J and reported Accuracy (ACC), Sensitivity (SENS), Precision (PREC), and Specificity (SPEC). Differences in AUC between the FC and SiFC models were evaluated using a two-sided paired permutation test on $\Delta \text{AUC} (B_{perm} = 5000),$ by randomly swapping the paired cross-validated predicted probabilities between models within each subject to form the null distribution.

We also performed partial least squares discriminant analysis (PLS-DA) to assess whether dynamic network measures distinguished weekly cocaine dosage within the CUD cohort. We categorized the $1–3$ g/week usage as low dosage and $4–6$ g/week as high dosage. The cutoff was selected to create clinically interpretable dosage groups while maintaining sufficient sample sizes for cross-validation. For each modality, the predictor set comprised $21$ dynamic network variables, including flexibility, recruitment, and integration across the seven Yeo-Krienen networks. FC and SiFC models were fitted separately on the same participants, and predictors were standardized within the cross-validation procedure. Model performance was evaluated using stratified $10$-fold cross-validation, with the number of latent variables selected based on cross-validated AUC. Classification performance was primarily summarized using AUC and balanced accuracy. Statistical significance was assessed using $2000$ label permutations. To compare the relative contribution of FC and SiFC, we summarized the aggregate Variable Importance in Projection (VIP) as the sum of VIP scores across the $21$ predictors for each modality, with error bars showing the foldwise standard deviation. After this block-level comparison, we examined predictor contribution within the SiFC model and interpreted variables with $\text{VIP} \ge 1$ as features with above-average contribution to the model \citep{chong2005performance}.

\section{Results}
\subsection{CUD is Associated with Higher Integration, Higher Recruitment, and Reduced Network Flexibility}

\begin{figure*}
    \centering
    \includegraphics[width=\textwidth]{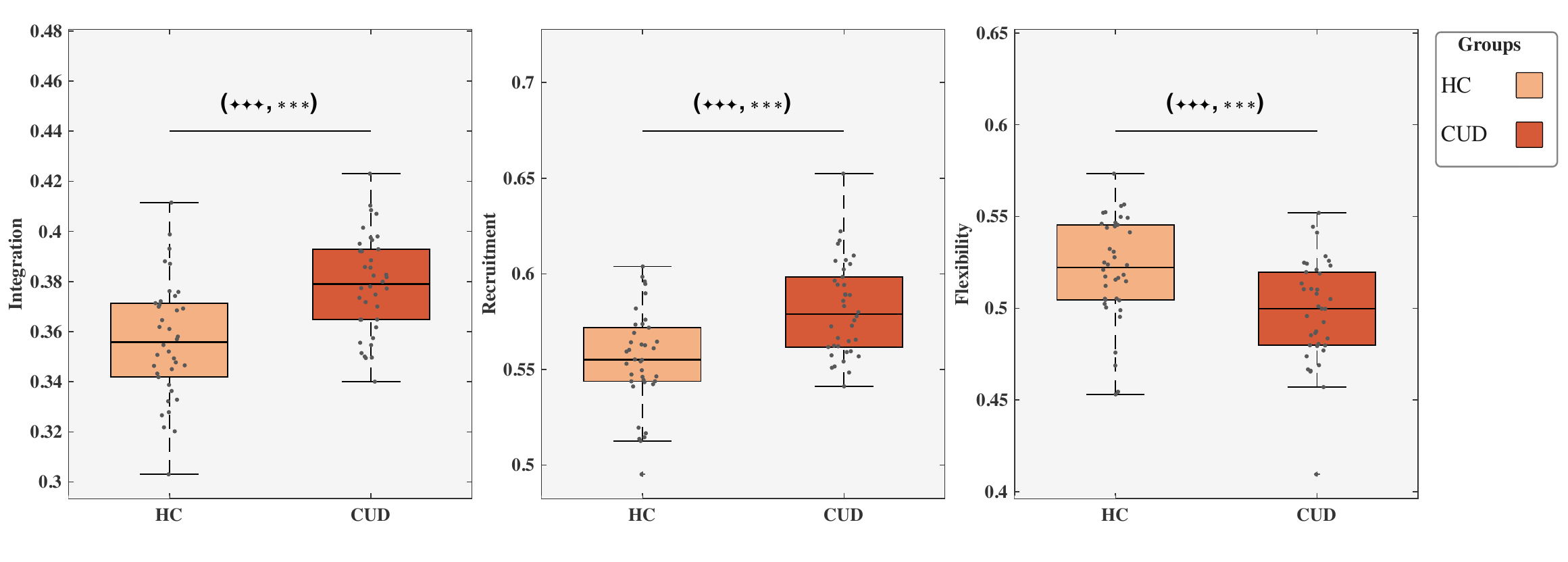}
    \caption{\textbf{Group differences in whole-brain network dynamics}. Comparison of global dynamic community metrics between healthy controls and cocaine use disorder participants ($N=38$ per group). The panels display the distribution of Integration (left), Recruitment (center), and Flexibility (right) averaged across the whole brain. Boxplots represent the median  and interquartile range (IQR). Individual data points represent single subjects. Statistical significance is reported above each comparison using the notation (uncorrected p-value, FDR q-value). Signs (\(\text{\FourStarSmall}\)) denote uncorrected $p$-values and ($*$) denote false discovery rate (FDR) corrected $q$-values, with significance levels defined as: \((\text{\FourStarSmall}\)$, \ast)$ for $< 0.05$,  \((\text{\FourStarSmall\ \FourStarSmall}\)$, \ast \ast)$ for $< 0.01$, and  \((\text{\FourStarSmall\ \FourStarSmall\ \FourStarSmall}\)$, \ast\ast  \ast)$ for $< 0.001$. }
    \label{fig:result1}
\end{figure*}

We characterized dynamic community organization at both whole-brain and individual network levels to identify neural systems impacted in cocaine use disorder. At the whole-brain level, CUD showed significantly higher dynamic integration ($p=0.0001$, $q=0.0001$) and recruitment ($p=0.0001$, $q=0.0002$), together with lower flexibility ($p=0.0005$, $q=0.0009$) relative to healthy controls (Fig.~\ref{fig:result1}). 

\begin{figure*}
    \centering
    \includegraphics[width=\textwidth]{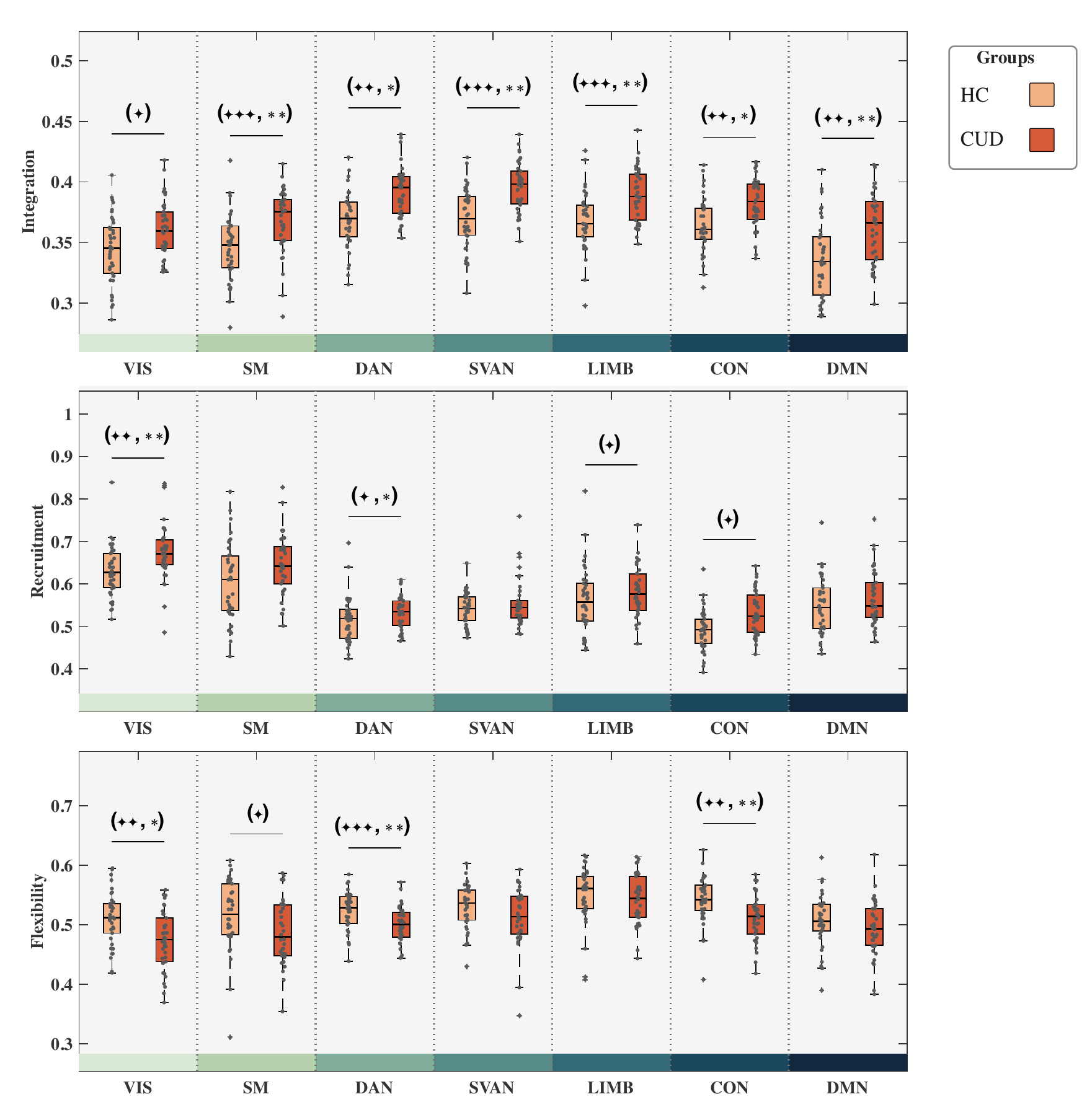}
    \caption{\textbf{Group differences in network-specific local dynamics}. Visualization of dynamic community metrics for the seven canonical resting-state networks including VIS (Visual), SM (Somatomotor), DAN (Dorsal Attention), SVAN (Salience/Ventral Attention), LIMB (Limbic), CON (Control), and DMN (Default Mode). The rows display Integration (top), Recruitment (middle), and Flexibility (bottom) across the 19 time windows. Boxplots show the median and interquartile range for healthy controls and cocaine use disorder. Statistical comparisons were performed using permutation testing with FDR correction for multiple comparisons. Significant differences are denoted by the notation (uncorrected, FDR-corrected) above the corresponding bars. Statistical significance is reported above each comparison using the notation (uncorrected p-value, FDR q-value). Signs (\(\text{\FourStarSmall}\)) denote uncorrected $p$-values and ($*$) denote false discovery rate (FDR) corrected $q$-values, with significance levels defined as: \((\text{\FourStarSmall}\)$, \ast)$ for $< 0.05$,  \((\text{\FourStarSmall\ \FourStarSmall}\)$, \ast \ast)$ for $< 0.01$, and  \((\text{\FourStarSmall\ \FourStarSmall\ \FourStarSmall}\)$, \ast\ast \ast)$ for $< 0.001$.}
     \label{fig:result2}
\end{figure*}

Network-level analyses indicated that these global differences reflected widespread but network-specific effects across canonical networks (Fig.~\ref{fig:result2}). Integration was elevated in CUD across all seven networks. These effects survived FDR correction in SM ($p=0.0001$, $q=0.0028$), DAN ($p=0.0036$, $q=0.0112$), SVAN ($p=0.0003$, $q=0.0028$), LIMB ($p=0.0004$, $q=0.0028$), CON ($p=0.0035$, $q=0.0112$), and DMN ($p=0.0014$, $q=0.0065$), whereas VIS ($p=0.0456$, $q=0.0851$) showed a nominal increase that did not survive correction. Recruitment increases were more selective, with FDR-significant effects in VIS ($p=0.0013$, $q=0.0065$) and DAN ($p=0.0188$, $q=0.0494$), whereas CON ($p=0.0299$, $q=0.0598$) and LIMB ($p=0.0294$, $q=0.0598$) showed nominal effects that did not survive correction. The SVAN ($p=0.9931$, $q=0.9931$) and DMN ($p=0.9328$, $q=0.9674$) showed no evidence of group differences. Flexibility was significantly reduced in CUD within VIS ($p=0.0043$, $q=0.0120$), DAN ($p=0.0004$, $q=0.0028$), and CON ($p=0.0017$, $q=0.0068$) networks after FDR correction, whereas no reliable differences were observed in SVAN ($p=0.2445$, $q=0.3423$), LIMB ($p=0.1469$, $q=0.2285$), SM ($p=0.0237$, $q=0.0553$), or DMN ($p=0.4060$, $q=0.5167$). Together, these findings indicate that individuals with CUD exhibit greater cross-network community co-assignment and stronger within-network community cohesion, accompanied by reduced temporal reconfiguration of community structure. These effects were most pronounced in visual, attentional, and control networks, suggesting a shift toward a more stable and less flexible pattern of large-scale brain network organization.

\begin{figure*}[ht]
    \centering
   \includegraphics[height=0.80\textheight, keepaspectratio]{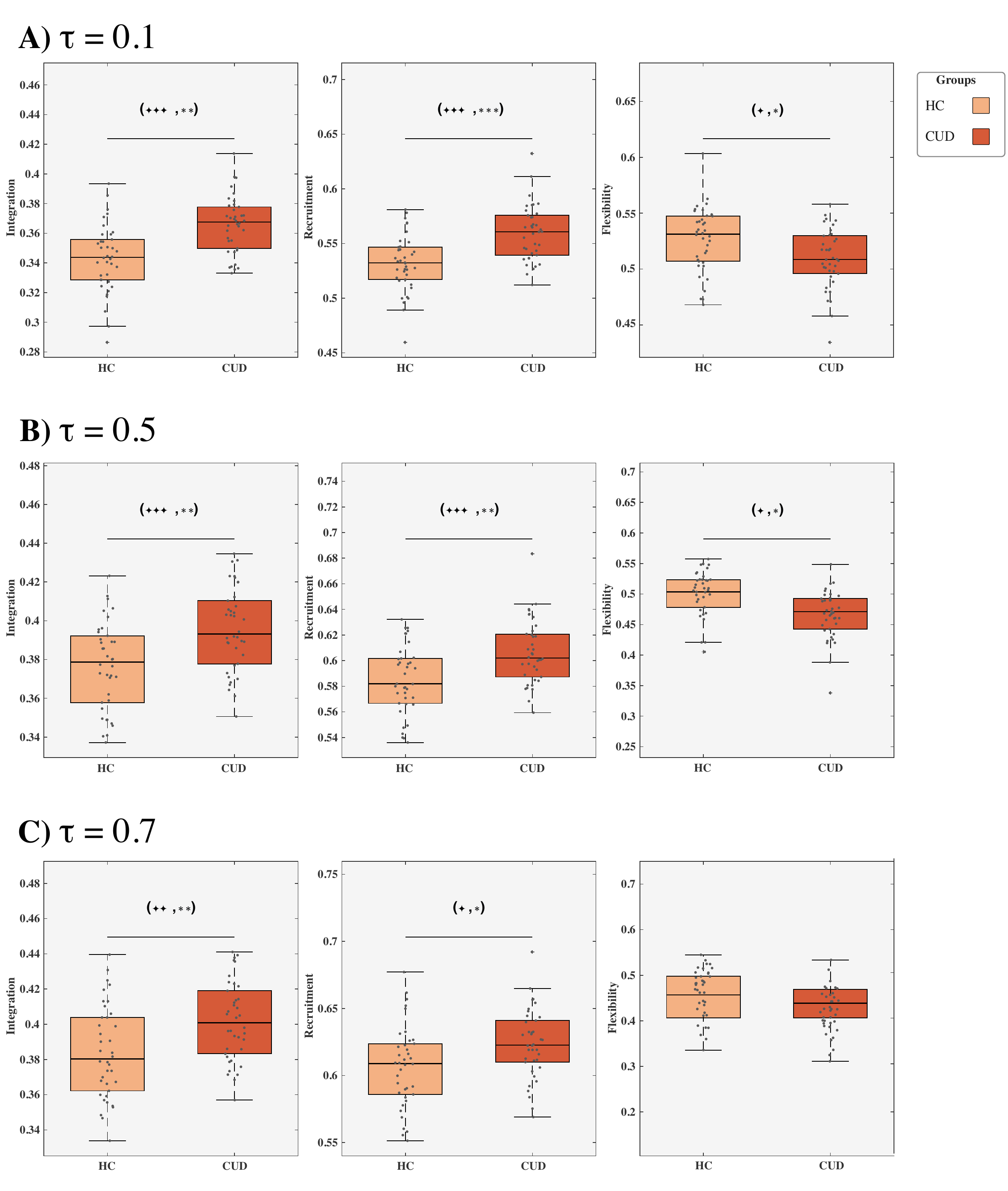}
    \caption{Sensitivity of whole-brain dynamic community measures to the structural prior strength $\tau$. Boxplots compare healthy controls and cocaine use disorder for Integration, Recruitment, and Flexibility computed from multilayer community structure after applying the structurally informed Laplacian-based spectral smoothing prior with (A) $\tau = 0.1$, (B) $\tau = 0.5$, and (C) $\tau = 0.7$.
Points show individual participants, boxes show the median and interquartile range.
Across $\tau$ values, CUD shows higher Integration and Recruitment and lower Flexibility, indicating that the primary group differences are robust to the choice of $\tau$.
Statistical significance is reported above each comparison using the notation (uncorrected p-value, FDR q-value). Signs (\(\text{\FourStarSmall}\)) denote uncorrected $p$-values and ($*$) denote false discovery rate (FDR) corrected $q$-values, with significance levels defined as: \((\text{\FourStarSmall}\)$, \ast)$ for $< 0.05$,  \((\text{\FourStarSmall\ \FourStarSmall}\)$, \ast \ast)$ for $< 0.01$, and  \((\text{\FourStarSmall\ \FourStarSmall\ \FourStarSmall}\)$, \ast\ast  \ast)$ for $< 0.001$.}
\label{fig:supp1}
\end{figure*}

\begin{figure*}
    \centering
   \includegraphics[height=0.70\textheight, keepaspectratio]{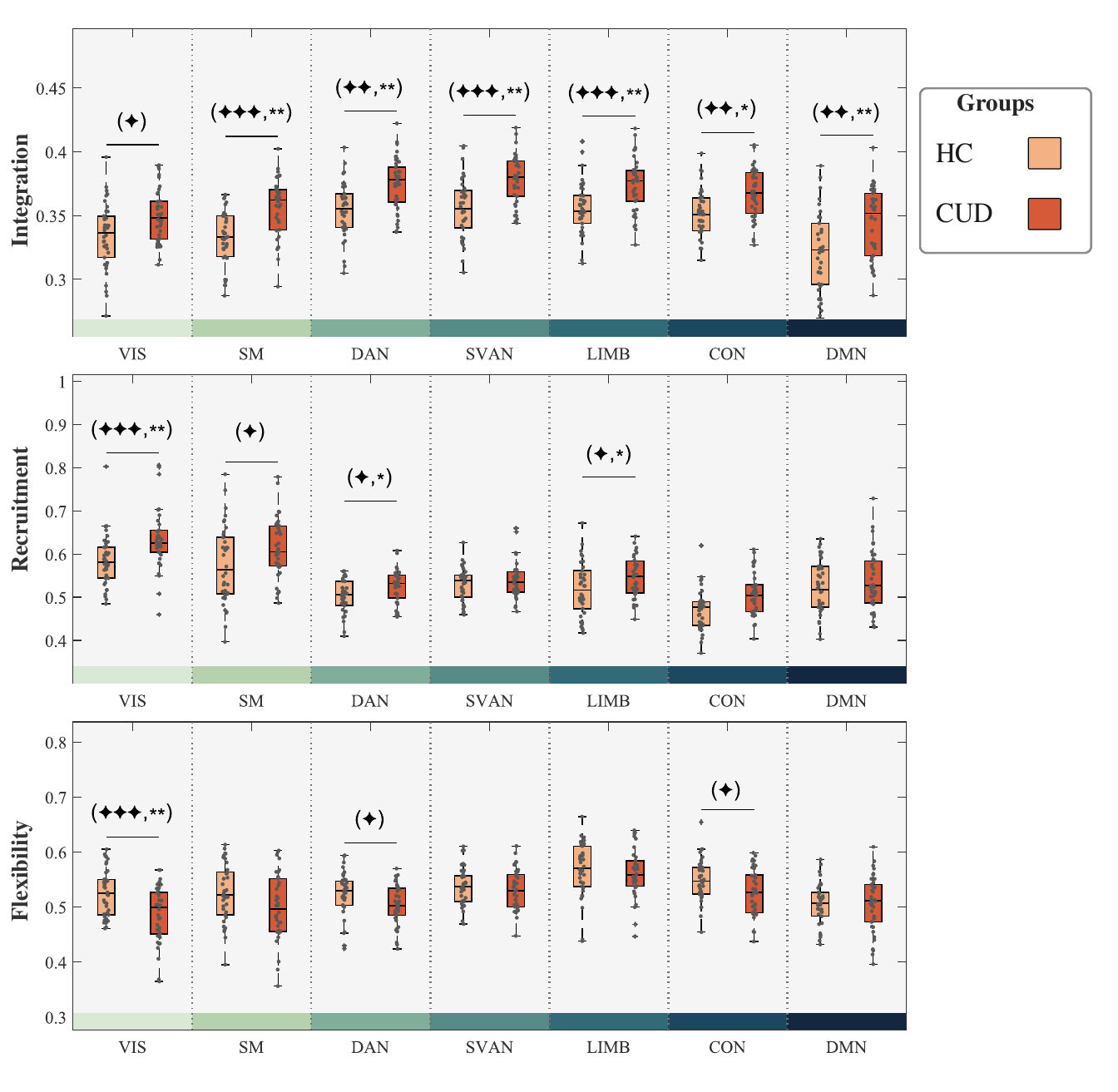}
    \caption{\textbf{Network-specific dynamic measures for $\tau=0.1$.} Visualization of dynamic community metrics for the seven canonical resting-state networks obtained using a structural prior smoothing parameter of $\tau=0.1$. The rows display Integration (top), Recruitment (middle), and Flexibility (bottom). Boxplots show the median and interquartile range for healthy controls and cocaine use disorder. Significant differences are denoted by the notation (uncorrected, FDR-corrected) above the corresponding bars. Statistical significance is reported above each comparison using the notation (uncorrected p-value, FDR q-value). Signs (\(\text{\FourStarSmall}\)) denote uncorrected $p$-values and ($*$) denote false discovery rate (FDR) corrected $q$-values, with significance levels defined as: \((\text{\FourStarSmall}\)$, \ast)$ for $< 0.05$,  \((\text{\FourStarSmall\ \FourStarSmall}\)$, \ast \ast)$ for $< 0.01$, and  \((\text{\FourStarSmall\ \FourStarSmall\ \FourStarSmall}\)$, \ast\ast \ast)$ for $< 0.001$.}
    \label{fig:supp2}
\end{figure*}

\begin{figure*}
    \centering
   \includegraphics[height=0.70\textheight, keepaspectratio]{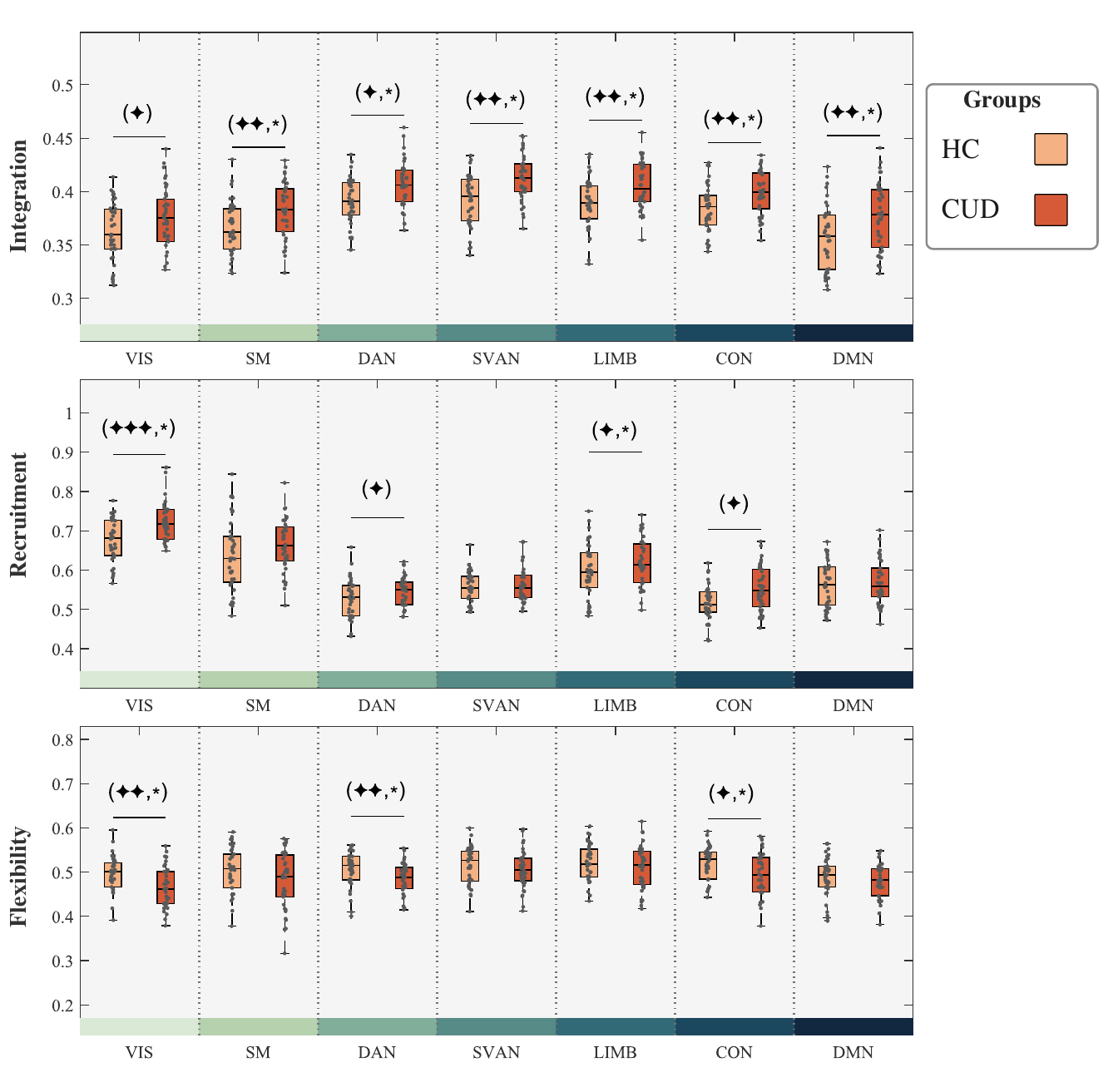}
    \caption{\textbf{Network-specific dynamic measures for $\tau=0.5$.} Visualization of dynamic community metrics for the seven canonical resting-state networks obtained using a structural prior smoothing parameter of $\tau=0.5$. The rows display Integration (top), Recruitment (middle), and Flexibility (bottom). Boxplots show the median and interquartile range for healthy controls and cocaine use disorder. Significant differences are denoted by the notation (uncorrected, FDR-corrected) above the corresponding bars. Statistical significance is reported above each comparison using the notation (uncorrected p-value, FDR q-value). Signs (\(\text{\FourStarSmall}\)) denote uncorrected $p$-values and ($*$) denote false discovery rate (FDR) corrected $q$-values, with significance levels defined as: \((\text{\FourStarSmall}\)$, \ast)$ for $< 0.05$,  \((\text{\FourStarSmall\ \FourStarSmall}\)$, \ast \ast)$ for $< 0.01$, and  \((\text{\FourStarSmall\ \FourStarSmall\ \FourStarSmall}\)$, \ast\ast \ast)$ for $< 0.001$.}
    \label{fig:supp3}
\end{figure*}

\begin{figure*}
    \centering
   \includegraphics[height=0.70\textheight, keepaspectratio]{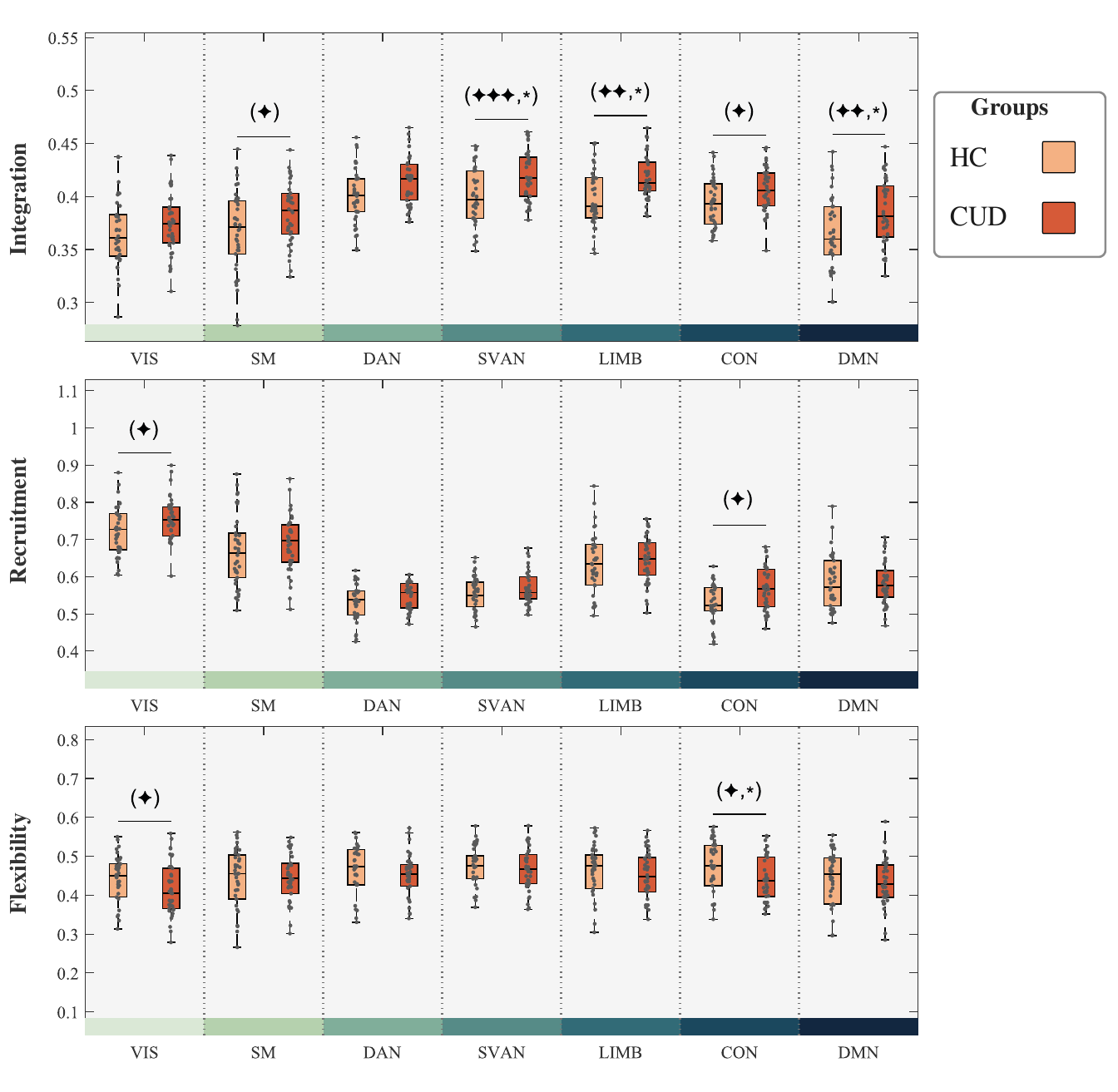}
    \caption{\textbf{Network-specific dynamic measures for $\tau=0.7$.} Visualization of dynamic community metrics for the seven canonical resting-state networks obtained using a structural prior smoothing parameter of $\tau=0.7$. The rows display Integration (top), Recruitment (middle), and Flexibility (bottom). Boxplots show the median and interquartile range for healthy controls and cocaine use disorder. Significant differences are denoted by the notation (uncorrected, FDR-corrected) above the corresponding bars. Statistical significance is reported above each comparison using the notation (uncorrected p-value, FDR q-value). Signs (\(\text{\FourStarSmall}\)) denote uncorrected $p$-values and ($*$) denote false discovery rate (FDR) corrected $q$-values, with significance levels defined as: \((\text{\FourStarSmall}\)$, \ast)$ for $< 0.05$,  \((\text{\FourStarSmall\ \FourStarSmall}\)$, \ast \ast)$ for $< 0.01$, and  \((\text{\FourStarSmall\ \FourStarSmall\ \FourStarSmall}\)$, \ast\ast \ast)$ for $< 0.001$.}
    \label{fig:supp4}
\end{figure*}

\begin{figure*}
    \centering
    \includegraphics[height=0.60\textwidth, keepaspectratio]{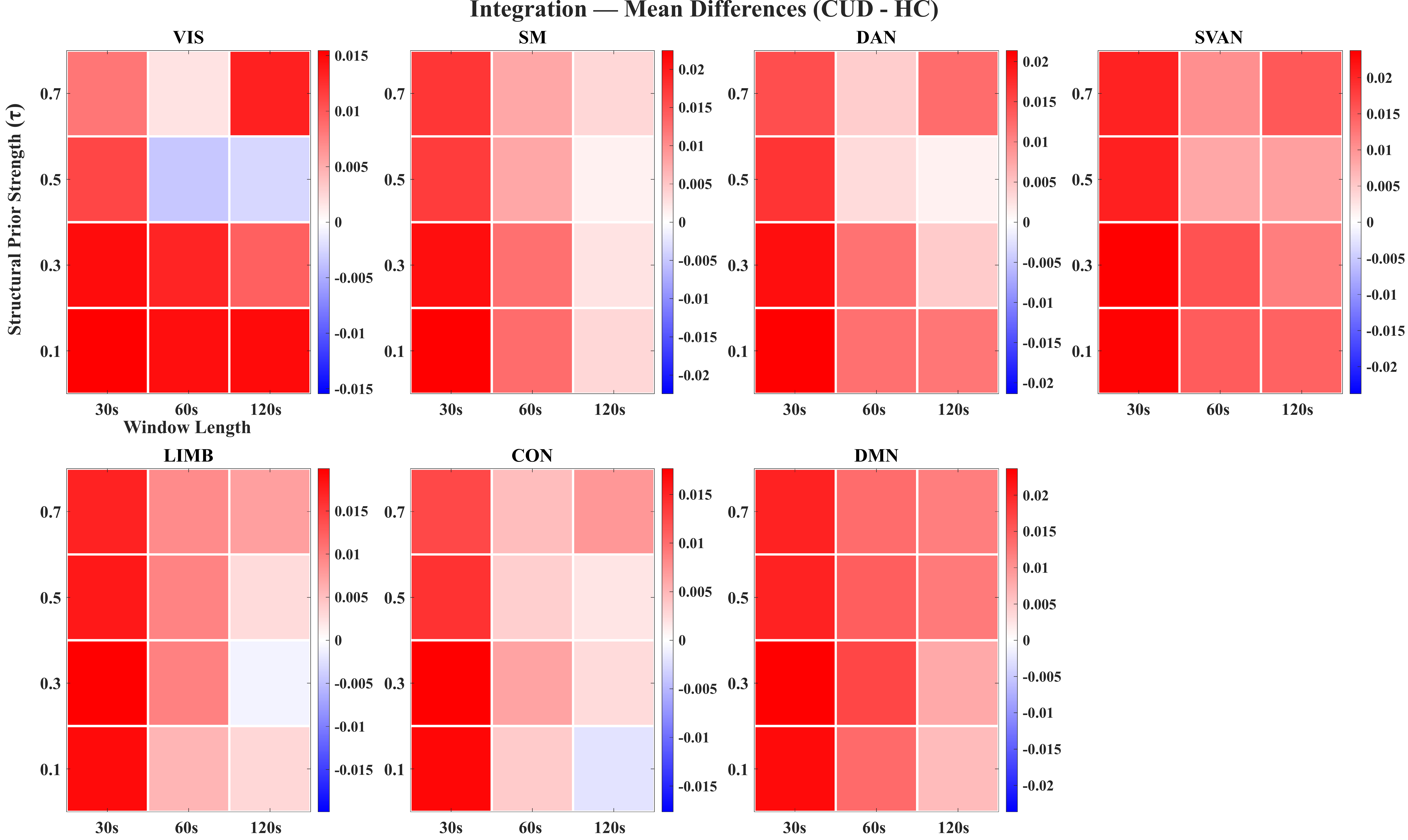}
   \par\vspace{1cm} 
    \includegraphics[height=0.60\textwidth, keepaspectratio]{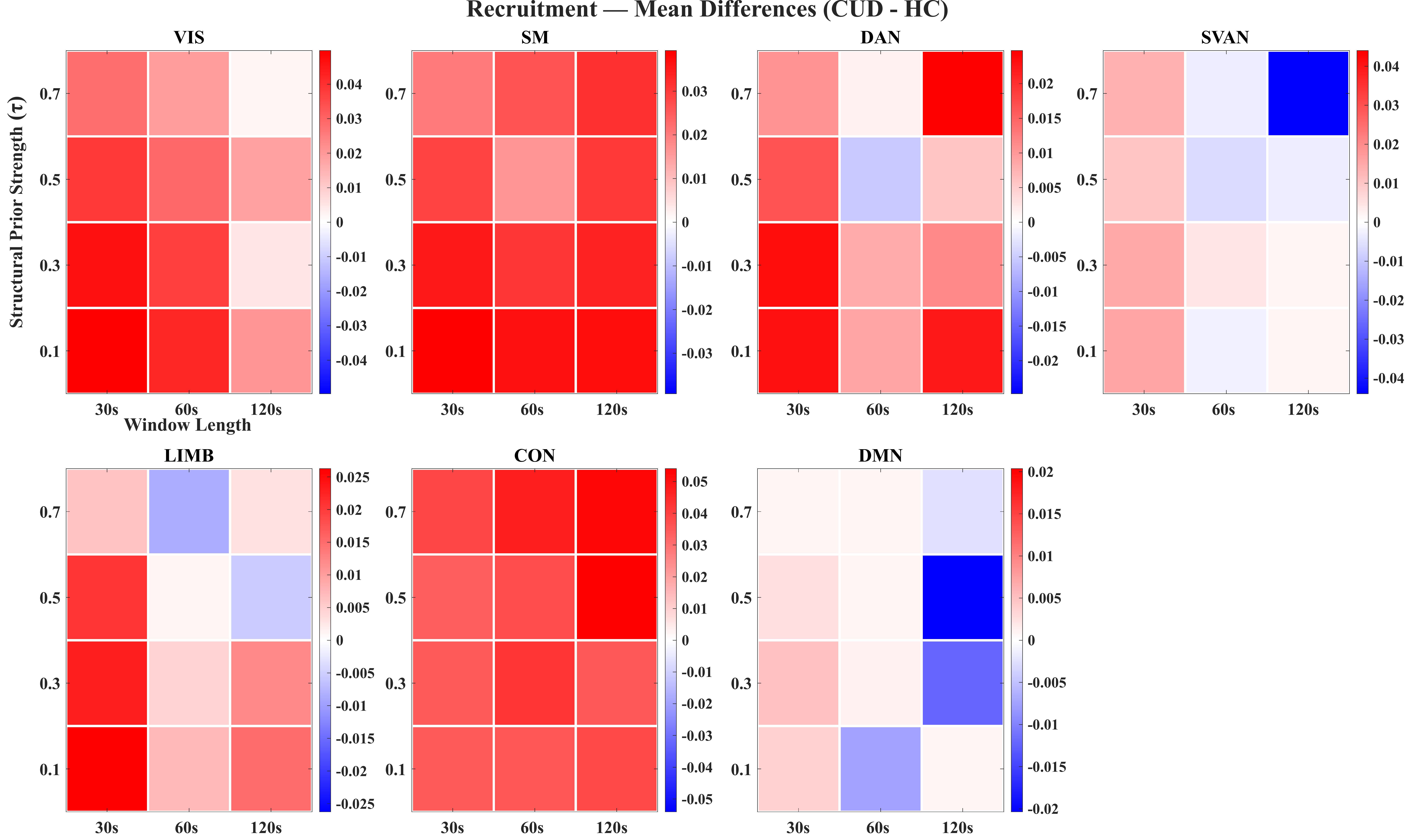}  
    \caption{Sensitivity analysis of network-level dynamic community measures across modeling parameters. Heatmaps show the mean group difference computed as CUD minus HC for Integration (top) and Recruitment (bottom) within each resting-state network (VIS, SM, DAN, SVAN, LIMB, CON, DMN). The x-axis varies the sliding-window length (30s, 60s, 120s) and the y-axis varies the structural prior strength \(\tau\) (0.1, 0.3, 0.5, 0.7). Red colors indicate higher metric values in CUD relative to HC ($CUD > HC$), while blue colors indicate higher values in HC relative to CUD ($HC > CUD$). The overall spatial pattern almost remains consistent across window lengths and \(\tau\), supporting robustness of the group differences to these parameter choices.}
\label{fig:supp5and6}
\end{figure*}

\begin{figure*}
    \centering
   \includegraphics[height=0.60\textwidth, keepaspectratio]{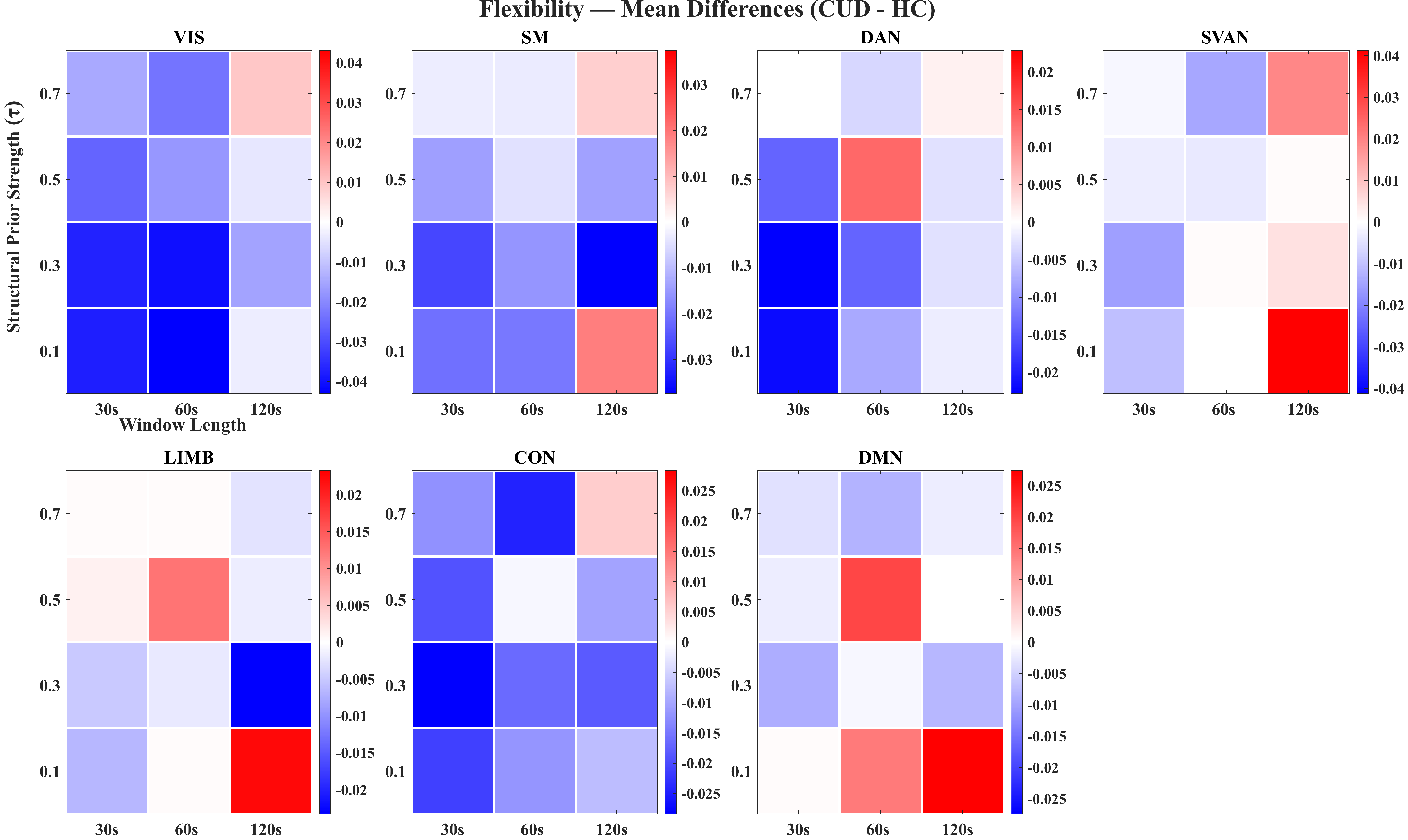}
  \caption{Sensitivity analysis of network-level flexibility across modeling parameters. Heatmaps show the mean group difference computed as CUD minus HC for Flexibility within each resting-state network (VIS, SM, DAN, SVAN, LIMB, CON, DMN). The x-axis varies the sliding-window length (30 s, 60 s, 120 s) and the y-axis varies the structural prior strength \(\tau\) (0.1, 0.3, 0.5, 0.7). Red colors indicate higher metric values in CUD relative to HC ($CUD > HC$), while blue colors indicate higher values in HC relative to CUD ($HC > CUD$). The overall pattern almost remains consistent across window lengths and \(\tau\), supporting robustness of the flexibility group differences to these parameter choices.}
    \label{fig:supp7}
\end{figure*}

To verify that our findings were not driven by the chosen parameter settings, we evaluated the robustness of the results across a range of structural prior strengths and temporal scales. First, we varied the structural smoothing parameter across $\tau \in \{0.1, 0.3, 0.5, 0.7\}$. At the whole-brain level, the direction of group differences was preserved across $\tau$, with higher integration and recruitment and lower flexibility in CUD, although the flexibility effect was attenuated at stronger regularization (Supplementary Fig.~\ref{fig:supp1}A-C). At the network level, integration showed the most consistent pattern across $\tau$, while recruitment and flexibility were more sensitive to structural regularization (Supplementary Figs.~\ref{fig:supp2}-\ref{fig:supp4}). We further repeated the analysis using $60$s and $120$s windows across different $\tau$ values (Supplementary Figs.~\ref{fig:supp5and6}-\ref{fig:supp7}). The overall pattern was broadly preserved across resting-state networks, although longer windows produced weaker and more heterogeneous network-level effects.

\subsection{SiFC Measures Enhance the Accuracy and Stability of CUD Classification}

\begin{figure*}
    \centering
    \includegraphics[width=\textwidth]{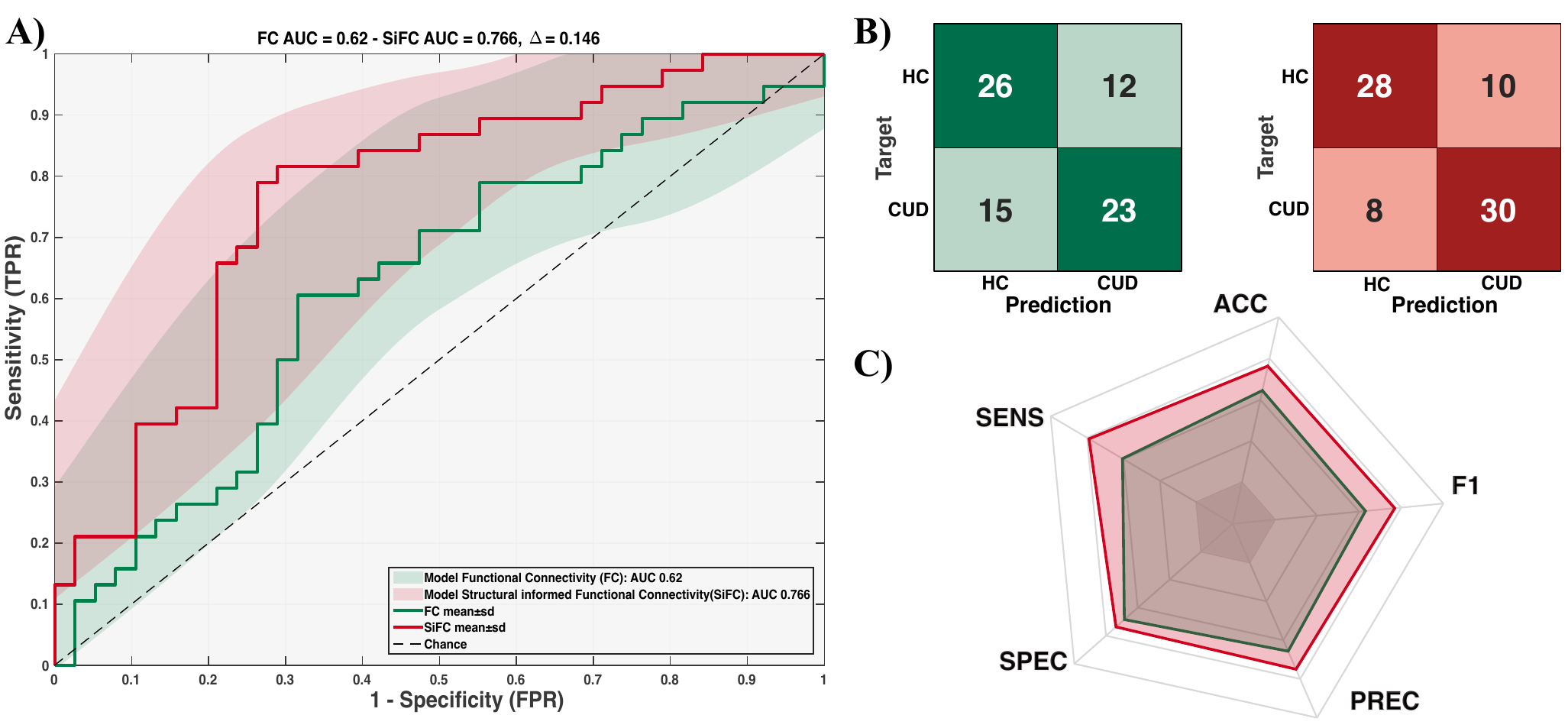}
    \caption{\textbf{Predictive power of functional connectivity features with and without structural priors for distinguishing cocaine use disorder from healthy controls}. \textbf{(A)} Receiver Operating Characteristic (ROC) curves displaying the Area Under the Curve (AUC) and model improvement ($\Delta$) based on logistic regression. Shaded regions indicate the standard deviation across $K=10$ cross-validation folds. The SiFC model consistently outperformed the FC baseline with $\Delta \mathrm{AUC}=0.146$. \textbf{(B)} Confusion matrices detailing prediction counts. The SiFC model increases true negatives (28 vs 26) and reduces false positives (10 vs 12) compared to the FC model. \textbf{(C)} Radar chart summarizing the performance of logistic regression models trained on FC and SiFC features across five metrics: Accuracy (ACC), F1-Score, Precision (PREC), Specificity (SPEC), and Sensitivity (SENS).}
\label{fig:result3}
\end{figure*}

We evaluated whether incorporating structural priors into functional connectivity improved the predictive power of network features for distinguishing individuals with CUD from healthy controls. We compared the structurally informed SiFC logistic regression model against the standard FC baseline using stratified 10-fold cross-validation. As shown in Fig.~\ref{fig:result3}, the SiFC framework demonstrated improved classification performance relative to the FC baseline. The SiFC model achieved an AUC of $0.766$ (95\% bootstrap CI: $0.651$ , $0.869$), compared with $0.620$ (95\% bootstrap CI: $0.488$ , $0.748$) for the FC model, yielding a $\Delta \mathrm{AUC} = 0.146$. This improvement was statistically significant under permutation testing ($p = 0.007$), indicating that the gain was unlikely to be driven by chance. Additionally, the smaller standard deviation across the ten cross-validation folds, represented by the shaded region in Fig.~\ref{fig:result3}A, suggests that incorporating structural constraints yields a more robust classification performance across different data partitions.

Confusion matrix analysis also revealed that the performance gain was driven by both improved detection of CUD and a reduction in False Positives (FP). Sensitivity, which reflects detection of CUD, increased from $0.605$ to $0.789$ with $30/38$ correctly identified, while specificity, which reflects correct rejection of controls, improved from $0.684$ to $0.737$ with $28/38$ correctly rejected. Consequently, the SiFC model achieved higher overall accuracy with $0.763$ versus $0.645$, precision $0.750$ versus $0.657$, and F1 score of $0.769$ versus $0.630$. This pattern is consistent with the structural prior acting as a regularizing constraint that reduces the influence of idiosyncratic functional fluctuations.

\subsection{SiFC Features Distinguish Higher Weekly Cocaine Use in CUD}

\begin{figure*}
       \centering\includegraphics[width=\textwidth]{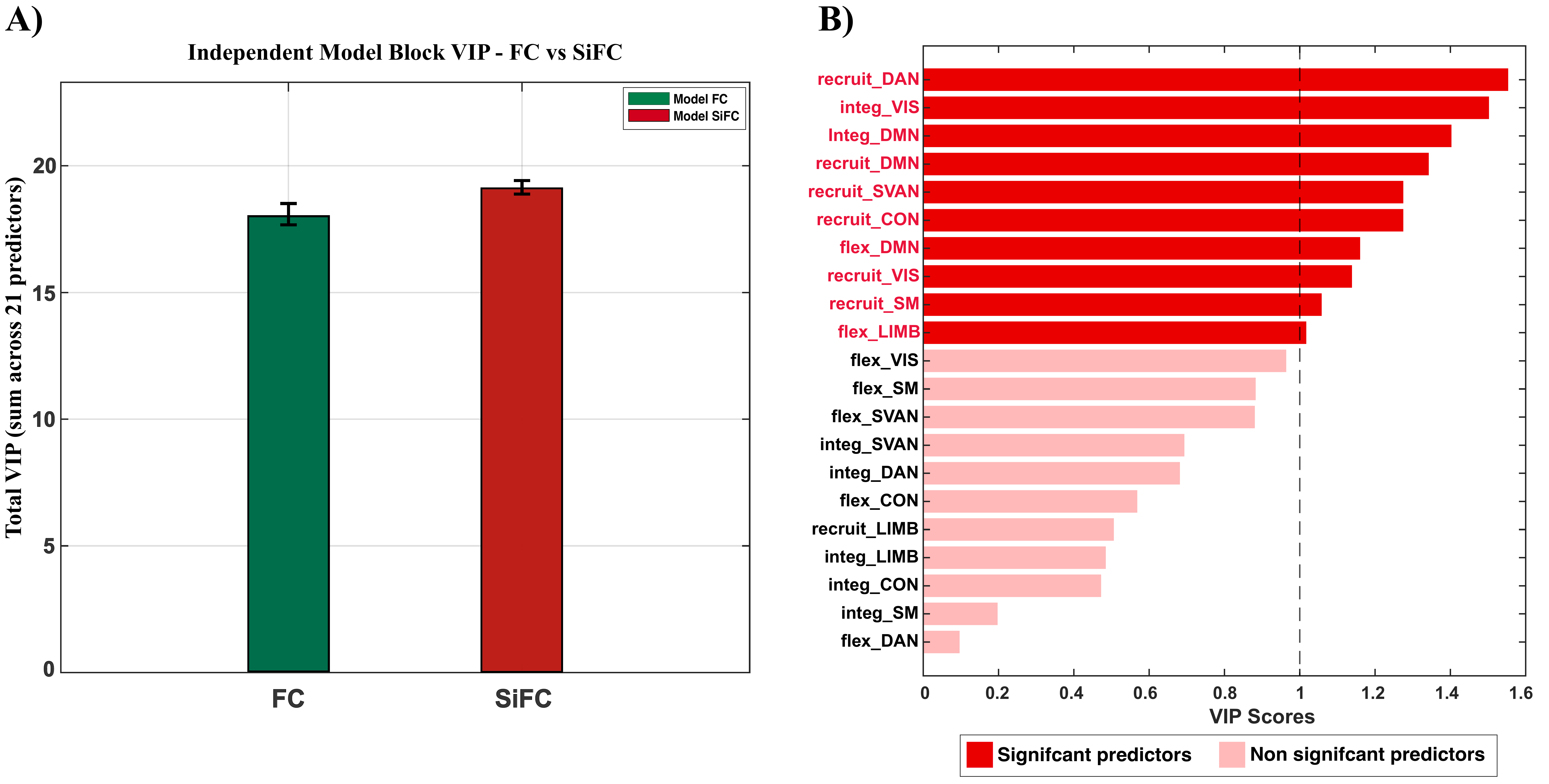}
\caption{\textbf{Predictive power of functional connectivity features with and without structural priors for detecting cocaine use dosage in CUD.} (A) Aggregate Variable Importance in Projection (VIP) comparison between independent FC and SiFC Partial Least Squares Discriminant Analysis (PLS-DA). The height of the bars represents the sum of VIP scores across the 21 dynamic network predictors, and error bars indicate the foldwise standard deviation. 
(B) Ranked VIP scores from the SiFC PLS-DA model. The dashed vertical line indicates the VIP = 1 threshold. Dark red bars indicate predictors with above-average contribution to the partial least square model, whereas light red bars indicate predictors below the threshold.}
       \label{fig:result4}
\end{figure*}

We also examined whether dynamic network measures distinguished high weekly cocaine use in the CUD cohort using partial least squares discriminant analysis (PLS-DA). As shown in Table~\ref{tab:plsda_fc_sifc_comparison}, the SiFC model outperformed the FC model across the main classification metrics. Specifically, SiFC achieved a higher AUC ($0.831$ vs. $0.781$) and balanced accuracy ($0.732$ vs.$0.659$). Both AUC and balanced accuracy were significantly greater than expected values under the permutation derived null distributions. Compared with the FC model, the SiFC model yielded lower permutation-based p-values for both AUC ($p=0.002$ vs. $0.019$) and balanced accuracy ($p=0.013$ vs. $0.059$), indicating stronger evidence for above-chance classification performance. Next, we compared the aggregate contribution of the two feature sets using total Variable Importance in Projection (VIP) scores. As shown in Fig.~\ref{fig:result4}A, SiFC had a higher total VIP than FC, with a total VIP of $19.22 \pm 0.33$ compared with $18.16 \pm 0.36$ for FC. Error bars reflect foldwise variability across the cross-validation folds, and total VIP was computed as the sum of VIP scores across the 21 dynamic predictors.

Lastly, we examined the contribution of SiFC predictors for identifying high weekly cocaine dosage (Fig.~\ref{fig:result4}B). The most influential SiFC features included integration within the VIS and DMN networks, recruitment within DAN, DMN, SVAN, CON, VIS, and SM, and flexibility within DMN and LIMB. Recruitment in DAN emerged as the strongest predictor, followed by integration in VIS and DMN and recruitment-related features spanning default mode, salience/ventral attention, control, visual, and somatomotor systems. Overall, these findings suggest that differences in weekly cocaine use are associated with a distributed structurally informed dynamic network signature involving cross-network community integration, within-network recruitment, and community flexibility across visual, default mode, attentional, control, somatomotor, and limbic systems.

Taken together, these findings suggest that incorporating structural information into dynamic functional connectivity could enhance the identification of clinically relevant network signatures of cocaine use. Compared with FC, the SiFC framework achieved stronger classification performance and greater aggregate VIP, indicating that structurally informed connectivity measures provide complementary information for characterizing individual differences in cocaine use intensity.

\begin{table}[!ht] 
\caption{Comparison of classification performance and aggregate predictor importance between FC and SiFC PLS-DA models.}
\label{tab:plsda_fc_sifc_comparison} 
\centering 
\begin{tabular}{lcc} \hline \textbf{Metric} & \textbf{FC} & \textbf{SiFC} \\ 
\hline 
AUC & 0.781 ($p=0.019$) & \textbf{0.831 ($p=0.002$)} \\ 
Balanced accuracy & 0.659 ($p=0.059$) & \textbf{0.732 ($p=0.013$)} \\ 
Total VIP $\pm$ SD & 18.16 $\pm$ 0.36 & \textbf{19.22 $\pm$ 0.33} \\ 
\hline 
\end{tabular}
\end{table}

\section{Discussion}
\subsection{Key Findings and Contribution}
We examined whole-brain dynamic community organization in cocaine use disorder using multilayer community detection applied to time-resolved and structurally informed functional connectivity. We summarized the resulting time-varying modular structure using three complementary measures: integration, recruitment, and flexibility. At the whole-brain level, the cocaine use disorder group showed higher integration, higher recruitment, and lower flexibility relative to healthy controls. This suggests a shift toward a more temporally stable and less reconfigurable pattern of large-scale brain network organization. This pattern is consistent with network-level rigidity, in which brain systems show greater persistence of community organization and reduced reconfiguration across time \citep{zhai2023disrupted,cong2024disrupted}. 

In this context, increased integration can be interpreted as a greater tendency for regions belonging to different canonical networks to participate in the same communities across time, suggesting that coordinated activity patterns extend across multiple large-scale functional systems. Increased recruitment further indicates greater persistence of within-system community cohesion across time, reflecting a stronger tendency for regions within a canonical network to remain grouped together across temporal layers. Along the same lines, reduced flexibility indicates that regions change their community assignments less frequently in CUD patients, suggesting reduced dynamic reconfiguration and greater temporal stability of community organization. 

At the network level, the effects were consistent with the whole-brain findings. Specifically, integration was higher across all resting-state networks, with significant effects in SM, DAN, SVAN, LIMB, CON, and DMN after FDR correction, while VIS showed a nominal increase. Recruitment increases were more pronounced in VIS and DAN, with nominal effects in LIMB and CON. Flexibility reductions were most pronounced in visual, attentional, and control networks, with a weaker nominal effect observed in the somatomotor network. This pattern aligns with addiction-related functions, including cue processing, attentional capture, habit-like responding, and impaired control \citep{ceceli2025impaired,le2025dynamic}. Because our analysis focused on positive functional interactions, increased integration reflects a greater tendency for regions from different canonical networks to be assigned to the same communities across time \citep{mattar2015functional}. Accordingly, the observed pattern suggests a more globally coordinated mode of functional organization in which VIS, SM, DAN, SVAN, LIMB, CON, and DMN regions increasingly participate in shared communities over time. This may provide a systems-level account for how drug cue and craving processes engage multiple large-scale systems and become embedded within ongoing resting-state dynamics \citep{zhai2023disrupted}.

The classification analyses further suggest that incorporating a subject-specific structural prior enhances the predictive power of dynamic network measures. Specifically, the structurally informed model achieved stronger discrimination between individuals with CUD and healthy controls, yielding a higher area under the curve than the FC model. This improvement was reflected across all classification metrics, including accuracy, F1 score, precision, sensitivity, and specificity. The confusion matrix similarly showed increases in true positives and true negatives, together with reductions in false positives and false negatives. Together, these findings indicate that structurally informed connectivity measures provide complementary information beyond functional dynamics alone, resulting in improved discrimination between groups.

The within-CUD classification analysis extends these findings by examining whether dynamic community measures are associated with differences in weekly cocaine use intensity. The SiFC model showed stronger classification performance than the FC model, with higher AUC and balanced accuracy, indicating that incorporating structural information enhances the ability to distinguish individuals with high weekly cocaine use. At the block level, SiFC features showed higher aggregate VIP and lower fold-to-fold variability than FC features, indicating a more stable set of influential predictors under cross-validation. The strongest contributors included integration in VIS and DMN, recruitment in CON, DAN, SVAN, DMN, SM, and VIS, and flexibility in DMN and LIMB. This feature profile suggests that higher weekly cocaine use is not characterized by alterations within a single network, but rather by a distributed dynamic signature. Specifically, higher weekly use was associated with greater integration of visual and default mode systems, together with increased recruitment of dorsal attention, default mode, salience/ventral attention, control, visual, and somatomotor networks. This interpretation is consistent with prior CUD studies showing that cocaine use dosage is associated with disrupted dynamic interactions among large-scale brain networks and distributed structural abnormalities \citep{zhai2023disrupted,yang2025brain}. 

\subsection{Limitations and Future Directions}
Despite the value of this work in providing a structurally informed framework for characterizing dynamic community organization in cocaine use disorder, several methodological limitations should be noted. First, the observed group differences in integration, recruitment, and flexibility cannot be interpreted as causal effects. It remains unclear whether the observed patterns reflect a pre-existing neurobiological vulnerability to cocaine use or a consequence of prolonged cocaine exposure. Future longitudinal research is needed to evaluate the predictive value of these dynamics and whether the observed alterations revert following sustained abstinence. 

Another consideration is that community detection was performed on positive functional connectivity networks. We adopted this approach because positive and negative functional connections play fundamentally different roles in community detection, with positive connections promoting the co-assignment of regions to the same community and negative connections favoring their separation into different communities. Consequently, the present findings characterize alterations in coordinated positive interactions among large-scale systems and do not directly address the contribution of anticorrelated network organization. Future work could compare positive-only and signed modularity formulations to determine how negative functional interactions contribute to dynamic community structure in cocaine use disorder.

Lastly, our sample size was modest ($N=76$), which may limit the generalizability of our findings. Future studies should replicate these results in larger and independent cohorts that include a broader range of clinical, demographic, behavioral, and cognitive characteristics, including variation in cocaine use severity, impulsivity, cognitive control, and treatment history. In addition, future work should examine the specificity of the identified network dynamics across substance use disorders and related psychopathology to determine whether the structurally informed measures capture CUD-specific neuroadaptations or reflect a more transdiagnostic alteration in large-scale networks.

\section{Conclusion}
In this work, we introduced a structurally informed dynamic connectivity framework to characterize large-scale brain network reconfiguration in cocaine use disorder. By incorporating subject-specific structural connectivity as an anatomical prior, we constrained time-resolved functional connectivity estimates and used multilayer community detection to quantify integration, recruitment, and flexibility across canonical brain networks. Our findings suggest that CUD is associated with a more rigid and less reconfigurable pattern of functional community organization, characterized by increased cross-network integration and recruitment together with reduced flexibility. These alterations were most evident in visual, attentional, and control networks, suggesting that addiction-related brain dynamics are characterized by greater participation of multiple systems within shared communities. Furthermore, the structurally informed network features not only improved discrimination between individuals with cocaine use disorder and healthy controls, but also captured differences in cocaine use intensity by distinguishing individuals with high versus low weekly cocaine use. The most discriminative features included integration of visual and default mode networks, recruitment across dorsal attention, salience/ventral attention, control, somatomotor, visual, and default mode networks, and flexibility within default mode and limbic systems. 

Taken together, these findings suggest that incorporating structural priors into dynamic functional connectivity estimation yields network measures that are both predictive and informative about network-level alterations associated with cocaine use disorder. Specifically, the structurally informed framework identified stronger cross-network positive community co-assignment and greater persistence of community structure in CUD, while also providing information related to individual differences in weekly cocaine use intensity. The results point to a promising direction for using structurally constrained dynamic network measures to identify stable signatures of cocaine use pathology. Such analyses could improve the characterization of individual differences in cocaine use intensity and support the development of network-based biomarkers for clinical stratification and treatment monitoring. Nonetheless, further studies are needed to validate these findings in larger and independent cohorts, examine their longitudinal stability, and determine whether these structurally informed dynamic signatures are specific to cocaine use disorder or reflect broader alterations across substance use disorders.

\section{Code and Data Availability}
All custom code used in this study is available at \href{https://github.com/3sigmalab/SiTMFC}{https://github.com/3sigmalab/SiTMFC}. The dataset used in this study is available at \href{https://openneuro.org/datasets/ds003346}{https://openneuro.org/datasets/ds003346}.
\section{Competing Interests}
No competing interest is declared.

\section{Author Contributions Statement}
SMR: Data Curation, Formal Analysis, Investigation, Methodology, Validation, Visualization, Writing (Original Draft), STH: Data Curation, Investigation, Visualization, TMT: Methodology, Investigation, MZ: Visualization, Investigation, AA: Visualization, Investigation, FZE: Writing (Review \& Editing), JAO: Writing (Review \& Editing), SK: Conceptualization, Investigation, Supervision, Writing (Review \& Editing).

\section{Acknowledgments}
The authors acknowledge the use of AI-assisted tools to improve the spelling, grammar, and readability of this manuscript.

\bibliographystyle{unsrtnat} 
\bibliography{reference}

\clearpage 
\onecolumn
\section{Supplementary Material} 

\end{document}